# A Multiclass Simulation-Based Dynamic Traffic Assignment Model for Mixed Traffic Flow of Connected and Autonomous Vehicles and Human-Driven Vehicles

Behzad Bamdad Mehrabani[a,*], Jakob Erdmann[b], Luca Sgambi[a], Seyedehsan Seyedabrishami[c], Maaike Snelder[d,e]

[a]Louvain Research Institute for Landscape, Architecture, Built Environment (LAB), Université Catholique de Louvain, Louvain-la-Neuve, Belgium
[b]German Aerospace Center (DLR), Berlin, Germany
[c]School of Civil Engineering, University of Sydney, Australia
[d]Transport & Planning Department, Delft University of Technology, Delft, Netherlands
[e]Netherlands Organization for Applied Scientific Research (TNO), The Hague, Netherlands
[*]Corresponding Author Email: behzad.bamdad@uclouvain.be

One of the potential capabilities of Connected and Autonomous Vehicles (CAVs) is that they can have different route choice behavior and driving behavior compared to human Driven Vehicles (HDVs). This will lead to mixed traffic flow with multiple classes of route choice behavior. Therefore, it is crucial to solve the multiclass Traffic Assignment Problem (TAP) in mixed traffic of CAVs and HDVs. Few studies have tried to solve this problem; however, most used analytical solutions, which are challenging to implement in real and large networks (especially in dynamic cases). Also, studies in implementing simulation-based methods have not considered all of CAVs' potential capabilities. On the other hand, several different (conflicting) assumptions are made about the CAV's route choice behavior in these studies. So, providing a tool that can solve the multiclass TAP of mixed traffic under different assumptions can help researchers to understand the impacts of CAVs better. To fill these gaps, this study provides an open-source solution framework of the multiclass simulation-based traffic assignment problem for mixed traffic of CAVs and HDVs. This model assumes that CAVs follow system optimal principles with rerouting capability, while HDVs follow user equilibrium principles. Moreover, this model can capture the impacts of CAVs on road capacity by considering distinct driving behavioral models in both micro and meso scales traffic simulation. This proposed model is tested in two case studies which shows that as the penetration rate of CAVs increases, the total travel time of all vehicles decreases.

# 1- Introduction and Background

The emergence of Connected and Autonomous Vehicles (CAVs) has revolutionized the transportation industry. The CAVs' technology has two main features: automation and connectivity. Automation refers to the ability to be driven by the combinations of human and machine decision-making and control systems. The driving automation ranges from 0 (fully manual) to 5 (fully autonomous). These levels of automation describe the sharing between humans and machines for controlling vehicles. The other feature (connectivity) can be considered as a feature that allows the vehicle to communicate with infrastructures (V2I), other vehicles (V2V), and pedestrians (V2P). Although these vehicles have not yet been fully integrated into the transportation network, much research has been conducted on how this type of vehicle affects the transportation network's performance. Conflicting debates exist on the impacts of CAVs on road networks; however, these vehicles are expected to have several advantages. From the perspective of traffic assignment, CAVs may have the following impacts on mixed traffic: 1- Due to their automation, CAVs' driving behavior will improve road capacity. This improved road capacity will lead to changes in link travel times, impacting the route choice of both CAVs and HDVs. 2- As a traffic management center may control CAVs, they may have full information about the status of the road network; hence they can follow different route choice principles compared to HDVs. For instance, they can follow System Optimal (SO) principles instead of User Equilibrium (UE). 3- Thanks to the connectivity feature of CAVs, they can respond to traffic congestion or disruption sooner than other types of vehicles by adjusting their routes (rerouting capability).

Although some studies, such as (Melson et al., 2018), have tried to solve the Traffic Assignment Problem (TAP) in the presence of CAVs, it should be taken into account that it may take a long time to have a 100% Penetration Rate (PR) of CAVs. Therefore, studying the above-mentioned impacts on different PR of CAVs (mixed traffic) is essential. To solve the TAP in mixed traffic, multiclass traffic assignment models are developed in the literature. These models investigate the route choices of heterogeneous users. This heterogeneity can be in terms of value of time, travel mode, travel disutility function, information quality, network topology, and different routing behavioral principles (Xie & Liu, 2022). The last group of heterogeneity causes (following other routing behavioral principles) can be formulated as a multiclass traffic equilibrium first addressed by Harker (Harker, 1988). However, the multiclass traffic equilibrium of CAVs and HDVs, following different routing behavioral principles, is a relatively new problem and needs more investigation. To this end, Table 1 provides the list of previous studies which implement or propose a framework for solving the multiclass TAP of mixed traffic of CAVs and HDVs. These studies solve the multiclass TAP in mixed traffic flow conditions either by considering the impacts of mixed traffic flow on road capacity or by considering different routing behavior between CAVs and HDVs. They can be examined from three perspectives: 1- The solution algorithm, 2- CAVs' scope of impact, and 3- CAV-related assumptions.

Regarding the solution algorithm, the proposed solution to the multiclass TAP in mixed traffic can be categorized into two groups: 1- Simulation-based models and 2- Analytical-based models (Mathematical Programming, Optimal Control Formulations, Variational Inequality-Based). Looking at Table 1, it is notable that the number of studies that implement the analytical-based methods (Aziz, 2019; Bagloee et al., 2017; Bahrami & Roorda, 2020; Chen et al., 2020; Guo et al., 2021; Li et al., 2018; Medina-Tapia & Robusté, 2019; Ngoduy et al., 2021; Sorani & Bekhor, 2018; G. Wang et al., 2020; J. Wang et al., 2019, 2021, 2022; Xie & Liu, 2022; F. Zhang et al., 2020; K. Zhang & Nie, 2018) is higher than the studies which implement the simulation-based techniques (Fakhrmoosavi et al., 2020; Hu et al., 2018; Liu et al., 2019; Mansourianfar et al., 2021;

Melson et al., 2018; Olia et al., 2016; Samimi Abianeh et al., 2020). Analytical models of multiclass TAP use analytical formulations to predict the propagation of traffic in a network (network loading). The Bureau of Public Roads (BPR) function is typically used in these studies to calculate link travel costs. Although the mathematical closed-form is available for the analytical solution algorithm, in practice, they cannot model certain phenomena (such as individual vehicles and vehicle interaction) in detail due to their macro-scale nature. On the other hand, applying analytical assignment problems to large-scale networks may be highly time-consuming and complex to solve (especially in dynamic cases) (Gawron, 1999). Simulation-based traffic assignment models use a traffic simulator to replicate the traffic flow dynamics and spatio-temporal interactions, which are based on micro/meso traffic flow simulation models (Saw et al., 2015). To capture CAVs' impacts on road capacity, these studies usually assume distinct driving behavior for CAVs (e.g. shorter reaction times (Mansourianfar et al., 2021); deterministic acceleration models (Fakhrmoosavi et al., 2020)). Compared to the analytical model, the simulation-based model appears to be more accurate because of its ability to explain traffic flow propagation in more detail (especially for CAVs).

Concerning the scope of the impacts, it can be argued that since these vehicles have not yet fully entered the transportation networks, each research has made its own assumptions. A review of studies shows that most of the existing literature examines the effects of CAVs, in solving the multiclass TAP of mixed traffic flow, either by their impacts on road capacity (driving behavior) (Aziz, 2019; Bahrami & Roorda, 2020; Ngoduy et al., 2021; F. Zhang et al., 2020) or by considering the fact that CAVs and HDVs follow different routing principles (either SO or UE) (Bagloee et al., 2017; Chen et al., 2020; Hu et al., 2018; Li et al., 2018; Medina-Tapia & Robusté, 2019; Olia et al., 2016; Samimi Abianeh et al., 2020; Sorani & Bekhor, 2018; J. Wang et al., 2019; Xie & Liu, 2022; K. Zhang & Nie, 2018). Although a few studies such as (Fakhrmoosavi et al., 2020; Liu et al., 2019; J. Wang et al., 2021, 2022) consider both CAVs' distinct route choice behavior and CAVs' impact on road capacity (driving behavior), they do not solve the multiclass equilibrium problem. They assumed that CAVs and HDVs follow the same routing behavioral principle (e.g., both classes follow UE principles). Few studies solve the multiclass equilibrium TAP (mixture of SO users and UE users) considering CAVs' impacts on road capacity (Guo et al., 2021; Mansourianfar et al., 2021; G. Wang et al., 2020). Also, less attention is paid to the effects of CAVs' rerouting capability in the multiclass equilibrium problem (Hu et al., 2018). For instance, Mansourianfar et al. 2021 (Mansourianfar et al., 2021) solved the multiclass equilibrium TAP by the simulation-based method. CAVs' impact on road capacity is modeled by applying modified parameters of a simplified car-following model in meso simulation. However, in Mansourianfar et al.'s work, CAVs' rerouting capability is not considered.

In terms of previous studies' assumptions about CAVs' route choice behavior, some researchers assume CAVs will follow the SO routines (Aziz, 2019; Bagloee et al., 2017; Chen et al., 2020; Guo et al., 2021; Li et al., 2018; Mansourianfar et al., 2021; Ngoduy et al., 2021; Sorani & Bekhor, 2018; G. Wang et al., 2020; K. Zhang & Nie, 2018), while others think CAVs will follow UE principles (Bahrami & Roorda, 2020; Samimi Abianeh et al., 2020; J. Wang et al., 2019, 2021; Xie & Liu, 2022; F. Zhang et al., 2020). Also, some studies model the deterministic route choice behavior of CAVs (G. Wang et al., 2020; J. Wang et al., 2019, 2021, 2022), and some others assume the stochastic route choice behavior for CAVs (Xie & Liu, 2022; F. Zhang et al., 2020).

To summarize, by reviewing previous articles on the solution framework of multiclass equilibrium TAP in mixed traffic flow, the following gaps can be expressed: 1- Most of the studies have used analytical solutions, which can be challenging in large dynamic real cases. 2- The solution frameworks capable of considering the impacts of CAVs on road capacity and the impacts of CAVs on route choice (multiclass equilibrium with the rerouting capability) are very rare. 3- There are no commonly accepted assumptions about the future route choice and driving behavior of CAVs.

Thus, this study fills these gaps by proposing a new open-source solution framework of the Multiclass Simulation-based Traffic Assignment Problem for the Mixed traffic flow of CAVs and HDVs (MS-TAP-M). This simulation-based solution framework considers the impact of CAVs on road capacity and the different route choice behavior of CAVs compared to HDVs (in micro and meso simulation scales). To address CAVs' impact on road capacity, modified parameters of car-following/lane-changing models (in microscale) and queuing model (in mesoscale) are utilized. Also, to distinguish between CAVs and HDVs in terms of route choice, it is assumed that HDVs follow UE while CAVs follow SO with rerouting capability. Since the solution framework is open-source, it is possible to model a variety of other assumptions on the driving behavior and route choice behavior of CAVs.

The remainder of this paper is organized as follows. The notations and abbreviations used in this paper are presented in section 2. Section 3 presents the methodology, followed by numerical results in section 4. The conclusions are presented in Section 5.

Table 1: Review of Previous Studies on Multiclass Traffic Assignment

| Author | Goal of the Study | TA Method | TA Approach | Software | CAVs impacts on capacity |
|---|---|---|---|---|---|
| Olia et al., 2016 | Assessing the impacts of CAVs | Simulation-Based (Micro) | DTA with rerouting capability (frequently rerouting for CAVs and infrequently rerouting for non-CAVs) | Paramics | No |
| Samimi Abianeh et al., 2020 | Evaluation of the impacts of CAVs on incidents | Simulation-Based (Micro) | DUE for CAVs and HDVs (considering rerouting capability for CAVs) | SUMO | No |
| Liu et al., 2019 | Evaluate the effect of route guidance under the CAV environment | Simulation-Based (Micro) | DUE for CAVs and HDVs (considering rerouting capability for CAVs) | PARAMICS | Yes |
| Hu et al., 2018 | Presenting DTA for Mixed Traffic | Simulation-Based (Meso) | Multiclass equilibrium (DSO and DUE) with rerouting capability | DynaTAIWAN | No |
| Fakhrmoosavi et al., 2020 | Observing the impacts of CAVs by adaptive fundamental diagrams | Simulation-Based (Meso) | DTA with rerouting capability for CAVs | DYNASMART-P | Yes |
| Mansourianfar et al., 2021 | Proposing a joint routing and pricing control scheme | Simulation-Based (Meso) | Multiclass equilibrium: DSO for CAVs (without rerouting) and DUE for HDVs | Aimsun | Yes |
| Bamdad Mehrabani et al., 2022 (**Current study**) | Proposing an open-source solution framework for multiclass TAP in mixed traffic flow | Simulation-Based (Micro and Meso) | Multiclass equilibrium: DSO for CAVs (with rerouting) and DUE for HDVs | SUMO | Yes |
| J. Wang et al., 2019 | Providing a solution algorithm for multiclass traffic assignment | Variational Inequality-Based | Mixed static traffic assignment: cross-nested logit for HDVs and UE for CAVs | N.A. | No |
| Li et al., 2018 | Providing a control Day-To-Day dynamical system to guide AVs | Variational Inequality-Based | Multiclass equilibrium: UE for HDVs and SO for CAVs | Matlab | No |
| K. Zhang & Nie, 2018 | Proposing a route control scheme | Variational Inequality-Based | Multiclass equilibrium with multiclass users: UE for HDVs and SO for CAVs | MATLAB | No |
| Xie & Liu, 2022 | Quantify the impacts of CAV on the vehicle market and route choices | Variational Inequality-Based | SUE with different perceived travel times for CAVs and HDVs | N.A. | No |
| G. Wang et al., 2020 | Providing a combined mode-route choice model for CAVs-HDVs | Variational Inequality-Based | Multiclass equilibrium: SUE for HDVs and SO for CAVs | N.A. | Yes |

Table 1: Review of Previous Studies on Multiclass Traffic Assignment

| Author | Goal of the Study | TA Method | TA Approach | Software | CAVs impacts on capacity |
|---|---|---|---|---|---|
| F. Zhang et al., 2020 | Studying the traffic equilibrium for mixed traffic flows of HDVs and CAVs | Variational Inequality-Based | SUE with different cost functions for CAVs and Non-CAVs | N.A | Yes |
| J. Wang et al., 2021 | Control the HDV using the CAV/toll lanes | Variational Inequality-Based | Cross nested logit with elastic jdemand for HDVs; UE with elastic demand for CAVs | N.A | Yes |
| J. Wang et al., 2022 | proposes a worst-case mixed traffic assignment model | Variational Inequality-Based | SUE for HDVs and UE for CAVs | N.A | Yes |
| Bagloee et al., 2017 | To highlight the benefit of the cooperative routing | Mathematical Programming | Multiclass equilibrium: UE for HDVs and SO for CAVs | GAMS | No |
| Sorani & Bekhor, 2018 | Evaluation in the presence of autonomous vehicles | Mathematical Programming | Multiclass equilibrium: UE for HDVs and SO for CAVs | N.A. | No |
| Medina-Tapia & Robusté, 2019 | Evaluating the effects of CAVs on a city road network | Mathematical Programming | Static assignment with a modified cost function for CAVs and HDVs | N.A. | No |
| Chen et al., 2020 | Develop a path-control scheme | Mathematical Programming | Multiclass equilibrium: SO for CAVs and UE for HDVs | CPLEX | No |
| Aziz, 2019 | Develop a mathematical program-based SO-DTA model for mixed traffic flow | Mathematical Programming | SO for both CAVs and HDVs | N.A. | Yes |
| Bahrami & Roorda, 2020 | Providing management policies for CAVs | Mathematical Programming | UE for CAVs and HDVs | N.A. | Yes |
| Guo et al., 2021 | Explore a system-level control mechanism of CAVs consisting of CAV and HDV | Mathematical Programming | Multiclass equilibrium: DSO for CAVs and DUE for HDVs | N.A. | Yes |
| Ngoduy et al., 2021 | Proposes a novel DSO formulation for the multiclass DTA problem | Mathematical Programming | DSO for both HDVs and CAVs | N.A. | Yes |

DSO: Dynamic System Optimal; DTA: Dynamic Traffic Assignment; DUE: Dynamic User Equilibrium; SUE: Stochastic User Equilibrium

## 2- Notations and Abbreviations

The used notations are listed in Table 2.

Table 2: Notations

| | |
|---|---|
| **Indices** | |
| $f$ | index for traffic flow |
| $i$ | index for iteration steps |
| $k$ | index for path |
| $j$ | index for vehicle |
| **Sets** | |
| $G(V, A)$ | traffic network |
| $D_H(R, S)$ | set of all HDVs ($J_H \in D_H$) |
| $D_C(R, S)$ | set of all CAVs ($J_C \in D_H$) |
| $J_H(r, s)$ | set of HDVs, travel from $r$ to $s$ ($j_h^{r-s} \in J_H$) |
| $J_C(r, s)$ | set of CAVs, travel from $r$ to $s$ ($j_c^{r-s} \in J_C$) |
| $A$ | set of links ($a \in A$) |
| $V$ | set of nodes ($v \in V$) |
| $R$ | set of origin nodes ($r \in R$) |
| $S$ | set of all destination nodes ($s \in S$) |
| $I$ | set of simulation iterations ($i \in I$) |
| $\delta_H$ | cardinality of set $D_H(R, S)$: number of HDVs O-D pairs |
| $\delta_C$ | cardinality of set $D_C(R, S)$: number of CAVs O-D pairs |
| $\pi_H$ | cardinality of set $J_H(r, s)$: number of HDVs travel from $r$ to $s$ |
| $\pi_C$ | cardinality of set $J_C(r, s)$: number of CAVs travel from $r$ to $s$ |
| $P_{j_h,i}^{r-s}$ | set of alternative paths for HDV $j_h^{r-s}$ in iteration $i$, travel from $r$ to $s$ |
| $P_{j_c,i}^{r-s}$ | set of alternative paths for CAV $j_c^{r-s}$ in iteration $i$, travel from $r$ to $s$ |
| **Variables, parameters, and elements** | |
| $c_a^i$ | travel time of link $a$ in iteration $i$ |
| $c_a^0$ | free flow travel time of link $a$ |
| $\bar{c}_a^i$ | marginal travel time of link $a$ in iteration $i$ |
| $C_k^i$ | travel time of path $k$ in iteration $i$ |
| $tt_{j_h,i}^{r-s}$ | experienced travel time of HDV $j_h$ in iteration $i$, travel from $r$ to $s$ |
| $\overline{tt}_{j_c,i}^{r-s}$ | experienced marginal travel time of CAV $j_c$ in iteration $i$, travel from $r$ to $s$ |
| $tt_{H,i}^{*,r-s}$ | least experienced travel time by HDVs in iteration $i$, travel from $r$ to $s$ |
| $\overline{tt}_{C,i}^{*,r-s}$ | least experienced marginal travel time by CAVs in iteration $i$, travel from $r$ to $s$ |
| $p_{j_h,i}^{r-s}$ | selected path for HDV $j_h^{r-s}$ in iteration $i$, travel from $r$ to $s$ |
| $p_{j_c,i}^{r-s}$ | selected path for CAV $j_c^{r-s}$ in iteration $i$, travel from $r$ to $s$ |
| $p_{j_h,i}^{*,r-s}$ | final selected path for HDV $j_h^{r-s}$ in iteration $i$, travel from $r$ to $s$ |
| $p_{j_c,i}^{*,r-s}$ | final selected path for CAV $j_c^{r-s}$ in iteration $i$, travel from $r$ to $s$ |
| $pr_{k,j_h}^i$ | probability of selecting path $k$ by HDV $j_h$ in iteration $i$ |
| $pr_{k,j_c}^i$ | probability of selecting path $k$ by CAV $j_c$ in iteration $i$ |
| $r_{j_c}^i$ | probability of rerouting by CAV $j_c$ during simulation in iteration $i$ |
| $\zeta$ | updating travel times of each link interval |

## 3- Methodology

In this section, the solution algorithm of MS-TAP-M is presented. The MS-TAP-M is defined as a dynamic TAP in which two vehicle classes exist. The first class of vehicles is CAV which follows the SO principle; the second class is HDV, which follows the UE principles. This framework can consider the rerouting behavior of CAVs while solving the MS-TAP-M. Also, it is possible to consider CAVs' impact on road capacity by setting some modified parameters of driving behavior on micro and meso scales.

The simulation-based solution of the dynamic TAP does not include any closed-form analytical solution and typically relies on an iterative procedure. An iterative scheme is employed to find the equilibrium solution in the single-class setting. These iterative methods start from an initial solution and update the path flow distribution for each iteration based on a path-swapping algorithm. The reassignment process of vehicles in each iteration confirms whether the algorithm is in a descent direction or not. In other words, the algorithm forces vehicles at each iteration to follow a more efficient path than the previous iteration. Also, at the end of each iteration, a convergence criterion (or error) is calculated to check the algorithm's termination. This approach was initially developed by Mahmassani and Peeta (Mahmassani & Peeta, 1993) and Peeta and Mahmassani (Peeta & Mahmassani, 1995). In this study, the framework provided by (Mahmassani & Peeta, 1993; Peeta & Mahmassani, 1995) has been developed to consider two classes of vehicles (CAVs and HDVs). Figure 1 illustrates the proposed solution framework for the MS-TAP-M. We name this solution framework "duaIterateMix". As shown in Figure 1, the framework consists of two parts: Path Selection Procedure (PSP) and Dynamic Network Loading (DNL). For the path selection procedure, Dijkstra's algorithm is used. For the DNL procedure, the Simulation of Urban Mobility (SUMO) (Lopez et al., 2018) is implemented. SUMO is an open-source, highly portable microscopic traffic simulation package that can handle large-scale networks. It also can conduct traffic simulation on the mesoscopic scale (DLR, 2021).

Consider $G(V,A)$ as the directed traffic network, which includes a set of links $A$ ($a \in A$) and a set of nodes $V$ ($v \in V$). $D_H(R,S)$ and $D_C(R,S)$ respectively represent the set of all HDVs and CAVs between all origin and destination pairs (travel demand). Where $R$ ($r \in R$) and $S$ ($s \in S$) denote the set of all origin nodes and the set of all destination notes, respectively. $J_H(r,s)$ ($J_H \in D_H$) is the set of HDVs travel from origin $r$ to destination $s$. $J_C(r,s)$ ($J_C \in D_H$) is the set of CAVs travel from origin $r$ to destination $s$. Hence, $j_h^{r-s}$ is an HDV that travels from origin $r$ to destination $s$; while $j_c^{r-s}$ is a CAV that travels from origin $r$ to destination $s$. The problem is stated as the assignment of $D_H(R,S)$ and $D_C(R,S)$ to $G(V,A)$. The multiclass equilibrium condition is computed by iteratively calculating the shortest routes and travel times. At each simulation iteration $i \in I$, first, a routing algorithm (Dijkstra) is applied to the road network to determine the set of alternatives paths, $P_{j_c,i}^{r-s}$ and $P_{j_h,i}^{r-s}$, for each vehicle $j_c^{r-s}$ and $j_h^{r-s}$. Thus, a new alternative path set is generated for each vehicle. For each CAV ($j_c^{r-s}$), the k-shortest paths are calculated using the previous simulation corresponding Marginal Travel Time (MTT), $\bar{c}_a^{i-1}$; while for each HDV ($j_h^{r-s}$) the k-shortest paths are calculated using the previous simulation's corresponding travel time, $c_a^{i-1}$. Next, a network loading model (Logit) is applied to the set of alternative paths, $P_{j_c,i}^{r-s}$ and $P_{j_h,i}^{r-s}$, to select a path, $p_{j_c,i}^{r-s}$ ($p_{j_c,i}^{r-s} \in P_{j_c,i}^{r-s}$); $p_{j_h,i}^{r-s}$ ($p_{j_h,i}^{r-s} \in P_{j_h,i}^{r-s}$). Then, a swapping algorithm is implemented to reassign a fraction of vehicles (within each class) at each iteration, ensuring the improvement of the selected path over iterations. Finally, the adjusted selected path (final path) of each vehicle type, $p_{j_c,i}^{*,r-s}$ and $p_{j_h,i}^{*,r-s}$, are determined and sent to SUMO to perform the traffic simulation. The traffic simulation can be performed either on microscopic or mesoscopic scales. On a microscale, the modified parameters of car-following/lane-changing models are implemented to consider the distinctions between CAVs' and HDVs' driving behavior. While on the mesoscale, the modified parameters of queuing model are used to distinguish between CAVs and HDVs. Besides, a predefined percentage of CAVs ($r_{j_c}^i$) reroute during simulation based on the current state of the traffic. At the end of the simulation, the current travel time of each link, $c_a^i$ will be calculated by SUMO. The travel time of each link is calculated according to the aggregated and averaged travel times of all vehicles on the corresponding link per defined interval ($\zeta$: 900 s). The travel times, written in this step $c_a^i$, are used as an input in the next iteration step. The total travel time (TTT) is minimized by performing such a process iteratively.

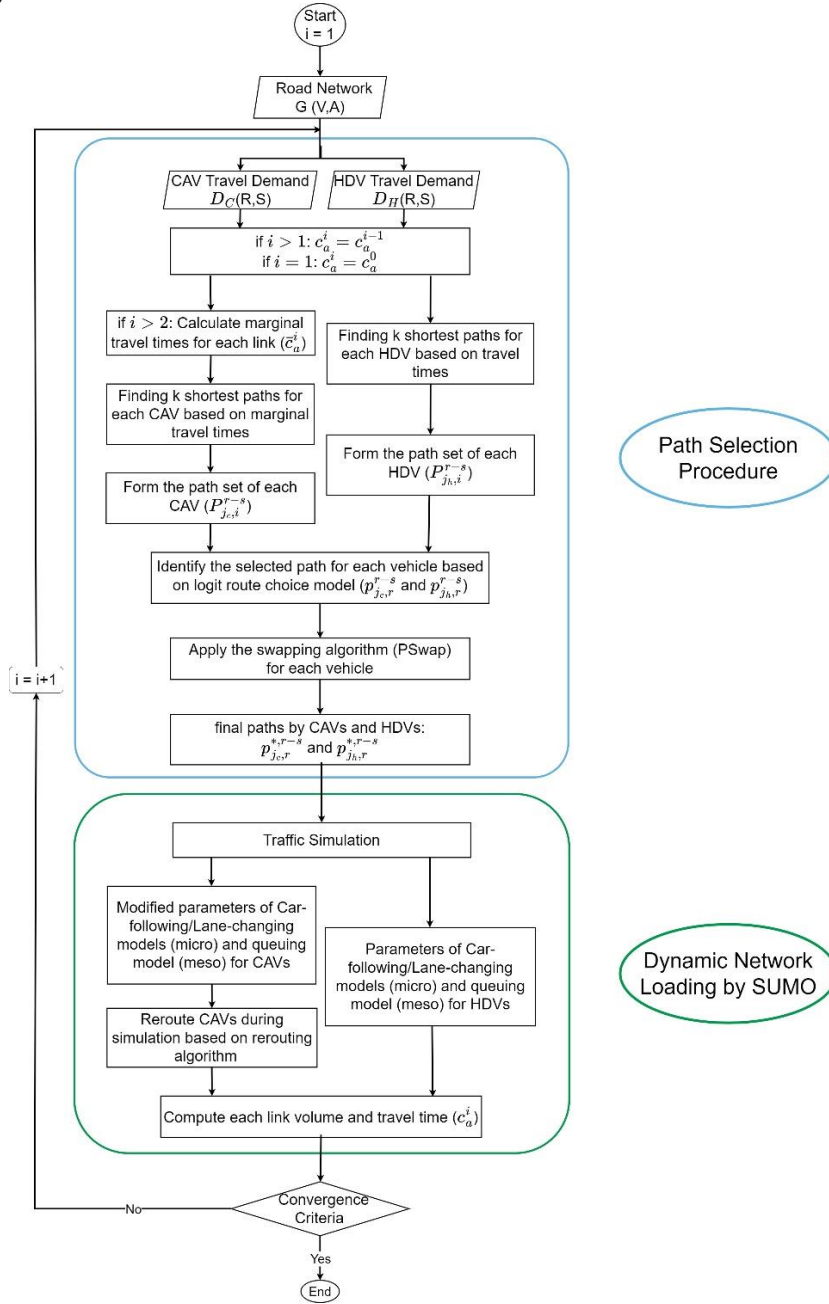

Figure 1: Framework of the MS-TAP-M

### 3-1- Path Selection Procedure

In the following sections, the steps of the path selection procedure are discussed in detail.

### 3-1-1- Calculation of Marginal Travel Times

In this framework, HDVs choose their path based on the previous iteration's link travel times (UE principle), while CAVs' path choice is based on the link MTT (SO principle). Two options are available to calculate the path MTT: 1- global approximation and 2- local approximation. The local approximation considers the path MTT as a summation of the corresponding link MTTs. In this study, a surrogate model of link MTT is implemented to calculate the local approximation of path MTT (Bamdad Mehrabani et al., 2022):

$$\bar{c}_a^i = c_a^{i-1} + f_a^{i-1} \frac{c_a^{i-1} - c_a^{i-2}}{f_a^{i-1} - f_a^{i-2}} \tag{1}$$

where $\bar{c}_a^i$ is the surrogate MTT of link $a$ at simulation step $i$; $c_a^{i-1}$ and $c_a^{i-2}$ are, respectively, the travel time (cost) of link $a$ at simulation steps $i-1$ and $i-2$; and $f_a^{i-1}$ and $f_a^{i-2}$ are, respectively, the traffic flow of link $a$ at simulation steps $i-1$ and $i-2$.

### 3-1-2- Route Choice Model

The route choice can be modeled as either deterministic or stochastic. In the deterministic approach, the all-or-nothing process is implemented (all of the flow between each O-D pair is assigned to the path with the shortest travel time). However, in the stochastic approach, the travelers are assigned to the routes based on probabilities (discrete choice models). The stochastic approach is chosen due to practical motivations: 1- the deterministic approach (all-or-nothing) is very sensitive to small changes in flow. For instance, for a simple network with only two parallel routes of equal length, the all-or-nothing procedure oscillates route choices since the drivers always choose the shortest route, which is the route with fewer cars. Hence, the solution is unstable even in this simple example with obvious equilibrium (each route is used by 50% of vehicles) (Gawron, 1999; Sheffi, 1985). 2- Given the multiple alternative routes with slightly different travel times, it may be reasonable to select a route other than the strictly shortest route (to avoid congestion on that route). The stochastic route choice models apply a scale parameter and converge to the optimum solution sooner than the deterministic approach.

The logit model is applied to each vehicle's set of alternative routes, $P_{c,i}^{r-s}$; $P_{h,i}^{r-s}$, in which the k-shortest paths for the subject vehicle are available. The travel times are considered as the cost for each alternative path. The (marginal) travel time of each path is equal to the sum of the (marginal) travel times of the corresponding links from the previous simulation. The logit model formulation is as follows

$$pr_{k,j_c}^i; \; pr_{k,j_h}^i = \frac{\exp(-\theta C_k^i)}{\sum_1^K \exp(-\theta C_K^i)} \tag{2}$$

$$C_k^i = \sum_{a \in A} \delta_{a,k}^i \, c_a^i \tag{3}$$

$$\delta_{a,p}^i = \begin{cases} 1 \text{ if link } a \text{ is on path } k \\ 0 \text{ otherwise} \end{cases} \tag{4}$$

where $pr_{k,j_c}^i$, $pr_{k,j_h}^i$ are the probability of selecting path $k$ by CAV $j_c$ and by HDV $j_h$ in iteration $i$; $C_k^i$ is the travel time (cost) of path $k$ in iteration $i$; and $\theta$ is the logit model scale parameter. Applying the scale parameter to the route choice procedure avoids the solution's oscillating and instability. It should be mentioned that although this study works on the stochastic solution of the traffic assignment problem, it is possible to reach the deterministic solution by the proposed algorithm.

*3-1-3- Swapping Algorithm*

The core idea of swapping algorithms is that not all vehicles should necessarily change their path in each iteration; instead, only a fraction of vehicles is in the reassignment process. Most previous studies apply the Method of Successive Average (MSA) as their swapping algorithm. However, this method has its own disadvantages (Sbayti et al., 2007). This study uses PSwap (Probabilistic Swapping) as the swapping algorithm, which is tested against MSA in the authors' previous work and shows better performance (Bamdad Mehrabani et al., 2022):

$$p_{j,i}^{*,r-s} = \begin{cases} p_{j,i}^{r-s} & if\ x \geq \rho_i \\ p_{j,i-1}^{*,r-s} & if\ x < \rho_i \end{cases} \quad (5)$$

Where $p_{j,i}^{*,r-s}$ is the final selected path by vehicle $j$ in iteration $i$; $p_{j,i}^{r-s}$ is the selected path by vehicle $j$ in iteration $i$ from the current logit model; $p_{j,i-1}^{*,r-s}$ the final selected path by vehicle $j$ in iteration $i-1$; $x$ is a random variable between 0 and 1; and $\rho_i$ is the sequence of step size in each iteration which can be considered as the probability of keeping the previous final selected path. In this study $\rho_i$ is predetermined $\rho_i = \frac{i}{\gamma}$; where $i$ is the iteration number, and $\gamma$ is a scale parameter. $\gamma$ is a real number that determines the speed of convergence. With a low value of $\gamma$, the speed of the convergence is fast, but few alternative paths are tested by each vehicle. On the other hand, with a high value of $\gamma$, the convergence speed is slow, while several alternative paths (available in the path set) will be tested by each vehicle. Accordingly, for stochastic assignments, it can be argued that higher values of $\gamma$ are preferable. However, waiting for a high number of iterations for large and medium-scale networks is computationally expensive. In this study, the value of $\gamma$ is set to 10 for the microscale and 50 for the mesoscale.

**3-2- Dynamic Network Loading**

Network loading is the process of assigning the O-D entries to the network for specific link travel times, described in the following sections. In this study, the dynamic network loading process is done by SUMO. This process can be performed on both the microscale and the mesoscale.

*3-2-1- Modeling CAVs and HDVs*

*3-2-1-1- Microsimulation*

Vehicles' movement is modeled by car-following and lane-changing models on the microscale in SUMO. In this study, the CAV modeling method in the microscale is the same as the work of Lu et al. (Lu et al., 2020) and Karbasi et al. (Karbasi et al., 2022). The core idea of modeling CAVs longitudinal movement is that they have the same car-following model as HDVs (Krauss car-following model); while some parameters are modified to simulate faster and safer behavior of CAVs. It is assumed that CAVs can always avoid collisions if the leader starts braking within the leader and follower's maximum acceleration bounds. The parameters modified for CAVs (level 5: Full Automation) (Krauss et al., 1997) in the Krauss car-following model are shown in Table 3.

Table 3: Parameters of the car-following model in SUMO

|     | Mingap (m) | Accel ($m/s^2$) | Decel ($m/s^2$) | Emergency Decel ($m/s^2$) | Sigma | Tau (s) |
|-----|------------|-----------------|-----------------|---------------------------|-------|---------|
| HDV | 2.5        | 2.6             | 4.5             | 8                         | 0.5   | 1.0     |
| CAV | 1.5        | 3.5             | 4.5             | 8                         | 0     | 0.9     |

Mingap: the offset to the leading vehicle when standing in a jam ($m$).
Accel: the acceleration ability of vehicles ($m/s^2$).
Decel: the deceleration ability of vehicles ($m/s^2$).
Emergency Decel: the maximum deceleration ability of vehicles in case of emergency ($m/s^2$).
Sigma: the driver imperfection (between 0 and 1).
Tau: the driver's desired (minimum) time headway (reaction time) ($s$).

The lateral movement of vehicles in SUMO traffic simulation is explained by LC2013 lane-changing model (Lopez et al., 2018). This model's parameters are modified to determine differences between CAVs and HDVs in lane-changing behavior. The most influential parameter in the LC2013 model is lcAssertive which describes the ''willingness to accept lower front and rear gaps on the target lane'' (DLR, 2021). With the higher values of lcAssertive, the vehicle's behavior toward shorter gaps is more aggressive, and the vehicle accepts lower gaps for lane-changing. In this research, the value of lcAssertive is set to 0.7 for CAVs and 1.3 for HDVs. For more information about the selection of the car-following/lane-changing parameters (for CAVs), please refer to (Karbasi et al., 2022).

*3-2-1-2- Mesosimulation*
The mesoscopic model of SUMO is based on the work of Eissfeldt (Eissfeldt, 2004). In this model, vehicles are placed in traffic queues which has a similar principle as the cell transmission model (Daganzo, 1995). Vehicles generally exit the queues in the order in which they entered them (first-in-first-out principle) (Amini et al., 2019). This model computes the time at which a vehicle travels from the queue based on the traffic state in the current and subsequent queue, the minimum travel time, and the stage of intersection (e.g. red, green, yellow). Based on the traffic state of each queue, four combinations exist between consecutive segments: 1- vehicle travel from free segment to free segment 2- vehicle travel from free segment to jammed segment 3- vehicle travel from jammed segment to free segment 4- vehicle travel from a jammed segment to jammed segment. The minimum headway between vehicles is then computed for each of the above combinations. The parameter $\tau$ is used to configure the minimum headways between vehicles (as a multiplier) for each of the four possible combinations.

Although some studies exist on modeling CAVs in mesoscale using a cell transmission model (Melson et al., 2018) or simplified car-following models (Mansourianfar et al., 2021), to the best of the authors' knowledge, no study exists on modeling CAVs with an Eissfeldt queuing model using SUMO which is different from the above-mentioned models. Therefore, to model CAVs on the mesoscale, the same approach of modeling CAVs on the microscale is implemented. The approach is to distinguish between CAVs and HDVs by applying different queuing model parameters (minimum headway: $\tau$) for each vehicle class. It is assumed that the $\tau$ parameter (headway) is lower for CAVs compared to HDVs as CAV can follow more efficiently between consecutive segments (Yu et al., 2021). The parameter in the mesoscale is calibrated for both CAVs and HDVs in a way that the mesoscopic model reveals the same Fundamental Diagram (FD) as the microsimulation. For various PRs of CAVs and HDVs, micro and meso simulations are performed, and the extracted FDs are compared with each other. For the start of the calibration process, the parameters of the micro simulation are set based on Table 3. The meso simulation parameter is extracted from the work of Presinger 2021 (Presinger, 2021). After calibration and comparing FDs of micro and meso scales, the value of $\tau$ is set to 1.06 for HDVs and 0.79 for CAVs. For more info on the parameters of mesoscopic simulation in SUMO and modeling CAVs in mesoscale, please refer to (Amini et al., 2019) and (Mansourianfar et al., 2021), respectively.

### 3-2-2- Rerouting

One of the CAVs' potential is that they are connected to each other and to a traffic management center (V2V and V2I); thus, they can send and receive real-time information about traffic conditions and each link's travel time. It is assumed that CAVs have the rerouting capability to address this feature in MS-TAP-M. The mechanism of rerouting in this study is that in the dynamic network loading step, a predefined percentage of CAVs ($r_{j_c}^i$) receive updated travel times both before insertion and periodically during movement in the network. According to this updated travel time, if they find a path shorter than the pre-selected path's travel time, they will change their path to the newly found shortest path.

### 3-3- Convergence Criterion

As no closed-form is available for the simulation-based solutions, it is impossible to prove the algorithm's convergence mathematically. All the convergence criteria only provide some point where the algorithm can be terminated. In UE condition, no driver can unilaterally reduce his/her travel time by shifting to another route. Previous studies on UE solutions typically calculate a gap function for the algorithm's convergence. However, a different convergence criterion should be considered in multiclass traffic conditions when there are two classes of vehicles (SO-seeking: CAVs and UE-seeking: HDVs). Similar to the study of Mansourianfar et al., 2021 (Mansourianfar et al., 2021), this study proposes a hybrid gap function for the algorithm's convergence. It is assumed that the multiclass traffic assignment condition is met when the travel time experienced by HDVs (UE-seeking) and the marginal travel time experienced by CAVs (SO-seeking) between the same OD pair are equal and minimal. For each class of vehicles, a relative gap is calculated, and the algorithm is considered as converged if the average of these two gaps becomes constant and less than $\varepsilon$.

$$Gap_1(i) = \frac{\sum_{H \in D_H} \left( \frac{\sum_{h \in H} tt_{j_h,i}^{r-s}}{\pi_H} - tt_{H,i}^{*,r-s} \right)}{\delta_H} \quad (6)$$

$$Gap_2(i) = \frac{\sum_{C \in D_C} \left( \frac{\sum_{c \in C} \overline{tt}_{j_c,i}^{r-s}}{\pi_C} - \overline{tt}_{C,i}^{*,r-s} \right)}{\delta_C} \quad (7)$$

$$Hybrid\ Gap(i) = \frac{Gap_1(i) + Gap_2(i)}{2} \quad (8)$$

Where $tt_{h,i}^{r-s}$ and $\overline{tt}_{c,i}^{r-s}$ are respectively experienced travel time of HDV $j_h$ in iteration $i$, travel from origin $r$ to destination $s$ and experienced marginal travel time of CAV $j_c$ in iteration $i$, travel from origin $r$ to destination $s$; $tt_{H,i}^{*,r-s}$ and $\overline{tt}_{C,i}^{*,r-s}$ are respectively least experienced travel time by HDVs in iteration $i$, travel from origin $r$ to destination $s$ and least experienced marginal travel time by CAVs in iteration $i$, travel from origin $r$ to destination $s$; $\pi_H$ and $\pi_C$ are the number of HDVs and CAVs traveling from origin $r$ to destination $s$, respectively; $\delta_H$ and $\delta_C$ are the number of HDVs and CAVs O-D pairs.

## 4- Numerical Results

The proposed algorithm was studied in two different networks (Figure 2): (a) medium-size abstract Random network and (b) large-size Sioux Falls network.

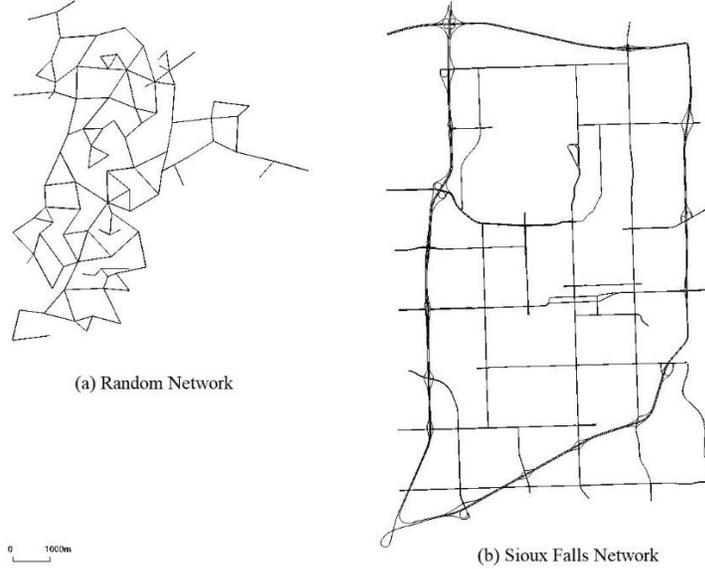

(a) Random Network

(b) Sioux Falls Network

Figure 2: Test Networks

A random network was implemented to evaluate the algorithm's performance in random unknown cases. Also, the algorithm is tested on the well-known Sioux Falls network, commonly used as a benchmark in the literature. In the following sections, the results of traffic simulations for the test networks are presented. The simulated scenarios for each of the networks are given in Table 4.

Table 4: Simulated Scenarios

| Network | Scenario | CAVs Penetration Rate | HDVs Penetration Rate | CAVs Rerouting Probability | Modeling CAVs driving behavior | CAVs routing principle | HDVs routing principle |
|---|---|---|---|---|---|---|---|
| Random Network | 1 | 0 | 100 | | | | |
| | 2 | 20 | 80 | | | | |
| | 3 | 40 | 60 | | | | |
| | 4 | 60 | 40 | | | | |
| | 5 | 80 | 20 | | | | |
| | 6 | 100 | 0 | 0.5 | yes | SO | UE |
| Sioux Falls Network | 1 | 0 | 100 | | | | |
| | 2 | 20 | 80 | | | | |
| | 3 | 40 | 60 | | | | |
| | 4 | 60 | 40 | | | | |
| | 5 | 80 | 20 | | | | |
| | 6 | 100 | 0 | | | | |

## 4-1- Random Network

A Random network was generated in SUMO (Figure 2 (a)). This network consists of 278 edges and 100 junctions. Each edge has a minimum length of 200 and a maximum length of 1000 meters. The number of lanes is either one or two for each edge. A random traffic demand of 7200 vehicles was generated for a one-hour simulation. These vehicles were randomly distributed to the network. As shown in Table 4, 6 different scenarios were simulated for this network. To investigate all potential capabilities of CAVs, it is assumed that they have different route choice and driving behavior in this case study. The distinct route choice behavior of CAVs is considered by taking that they follow SO and have the rerouting capability, while the particular driving behavior is addressed by modifying the relevant car-following model's parameters, as discussed in the previous sections. The microsimulation is performed, which converged after 10 iterations. TTT of all vehicles in each iteration for each scenario is displayed in Figure 3.

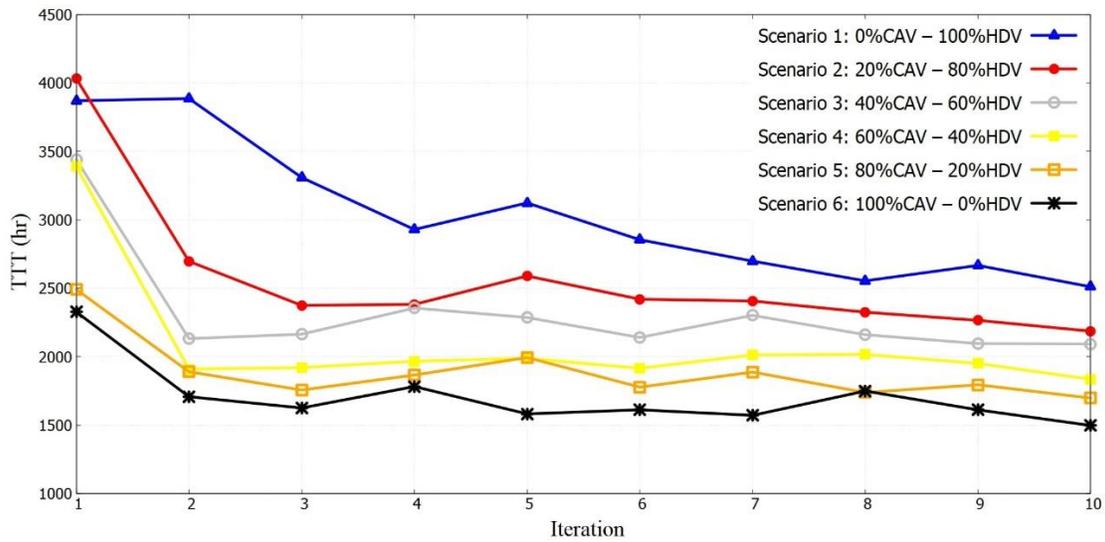

Figure 3: Convergence patterns for Random network

The results of the simulations for the Random network are given in Table 5. A glance at Table 5 reveals that as the PR of CAVs increases, the TTT of vehicles decreases. The percentage of TTT reduction in comparison to scenario 1 (0% PR of CAVs) varies from 12.9% (for 10% PR of CAVs) to 48.9% (for 100% PR of CAVs). This shows that if all vehicles have the full automation driving mode (100% PR of CAVs) and follow the SO routines with 50% of the rerouting probability, the TTT is reduced by 48.9% compared to the 0% PR of CAVs. In addition, we are observing the increasing trend of average speed and the decreasing trend of average distance traveled by vehicles as the PR of CAVs increases.

Table 5: Simulation Results for Random Network

| Scenario | Hybrid Gap | Total Travel Time (hr) | Average Speed (km/h) | Average Distance Travelled (km) | TTT improvement (%) |
|---|---|---|---|---|---|
| 1 | 12.49 | 2509.88 | 29.2 | 6.6 | |
| 2 | 10.53 | 2185.54 | 31 | 6.6 | 12.9 |
| 3 | 9.86 | 2092.5 | 32 | 6.6 | 16.6 |
| 4 | 8.12 | 1834.36 | 34 | 6.4 | 26.9 |
| 5 | 7.44 | 1696.94 | 35.4 | 6.5 | 32.3 |
| 6 | 5.91 | 1496.72 | 37.5 | 6.4 | 48.9 |

Figure 4 shows traffic volume (veh/hr) in the Random network in different PRs of CAVs. This figure depicts that as the PRs of CAVs increases, the number of medium-volume and high-volume links increases. This fact can be justified based on two reason. First, as the number of CAVs increases, links' capacity increases, so they can service more vehicles. As a result, the volume of links which are part of vehicles' shortest paths increases. The second reason can be that as CAVs follow SO principles, they are distributed throughout the entire network and use the spare capacity. With high PR of CAVs, the volume of alternatives routes for O-D pair increases. As a consequence, the number of blue links in 100% PR of CAVs is more than 0% PR of CAVs.

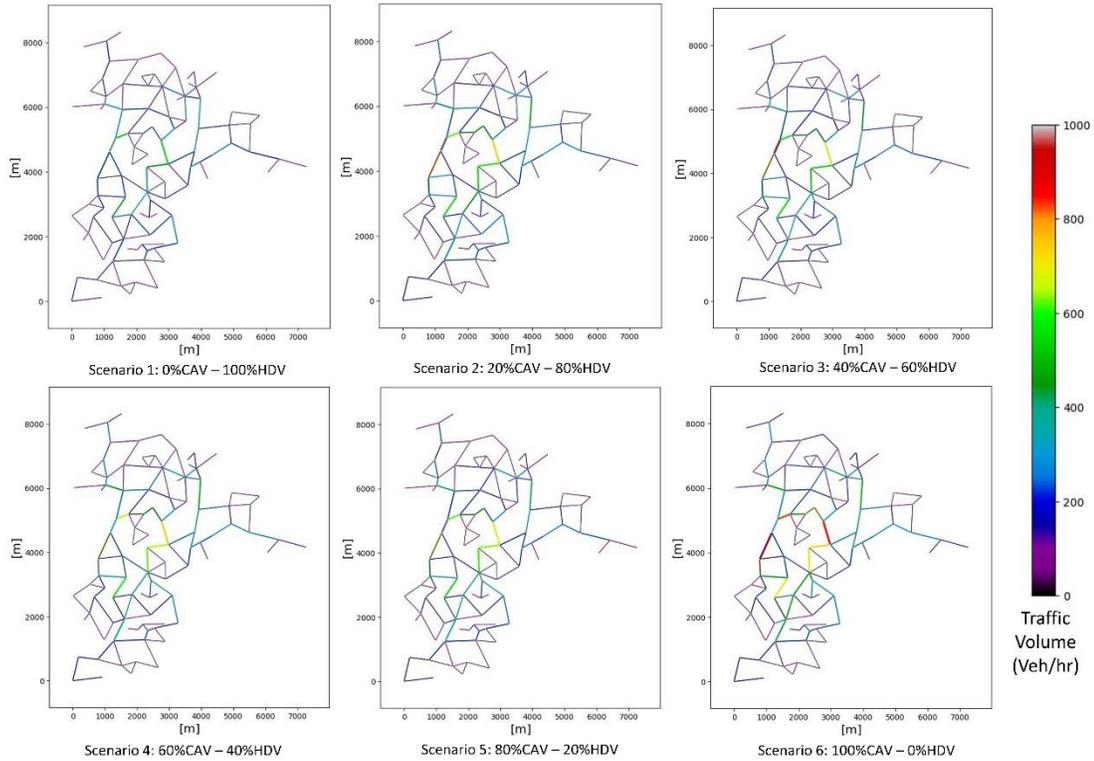

Figure 4: Traffic Volume of Random Network in different Penetration Rate of CAVs

The scenarios that have been studied so far for the random network are the scenarios in which we witness a reduction in TTT due to the combined impact of CAVs' different driving and route choice behavior. To have an in-depth view of the impacts of CAVs on TTT, several other scenarios have been investigated. Twelve new scenarios are simulated which either consider CAVs' different route choice behavior or CAVs different driving behavior (2*6 scenario). These new simulations have been compared with the results of the scenarios that have been done so far. The results of this comparison are illustrated in Figure 5. In this figure, the TTT for three different categories of scenarios is shown. These categories include 1- scenarios in which only CAVs' different route choice behavior is modeled (Route choice impact) 2-scenarios in which only CAVs' different driving behavior is modeled (Capacity impact) 3- scenarios in which both CAVs' different route choice and driving behavior is considered (Combined impact). This figure also displays the amount of improvement in TTT for each of the scenarios mentioned above. By having a glance at Figure 5, it can be recognized that when CAVs have solely different route choice behavior in comparison to HDVs, they have the least improvement in TTT. However, when they have both different route choices and driving behavior, we observe the best improvement in TTT.

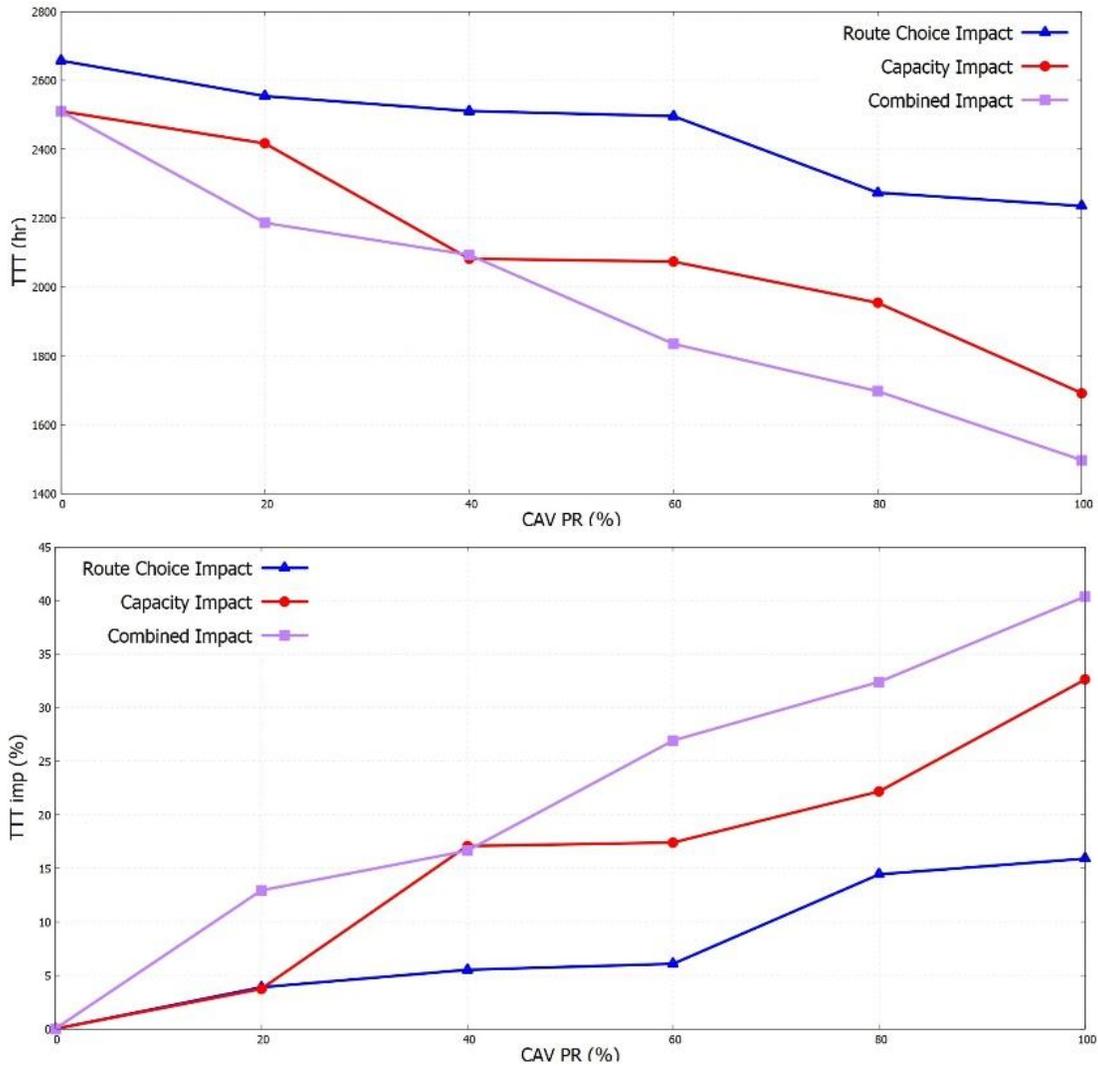

Figure 5: Impact of CAVs on TTT for Different Setting of Route Choice and Driving Behavior (Random Network)

### *4-2- Sioux Falls Network*

The second case study in this article is the Sioux Falls network (Figure 2 (b)). The total number of simulated vehicles was 36000, distributed on different origins and destinations based on the demand pattern of LeBlanc's study (LeBlanc et al., 1975). The mesoscopic simulation is used for this case study. Six scenarios were analyzed, and the convergence pattern for all scenarios (scenarios 1 to 6) is shown in Figure 6. For initial investigations, it is assumed that CAVs have different driving and route choice behavior. CAVs follow SO with rerouting capability, while HDVs follow UE. Also, CAVs have lower headway in meso simulation.

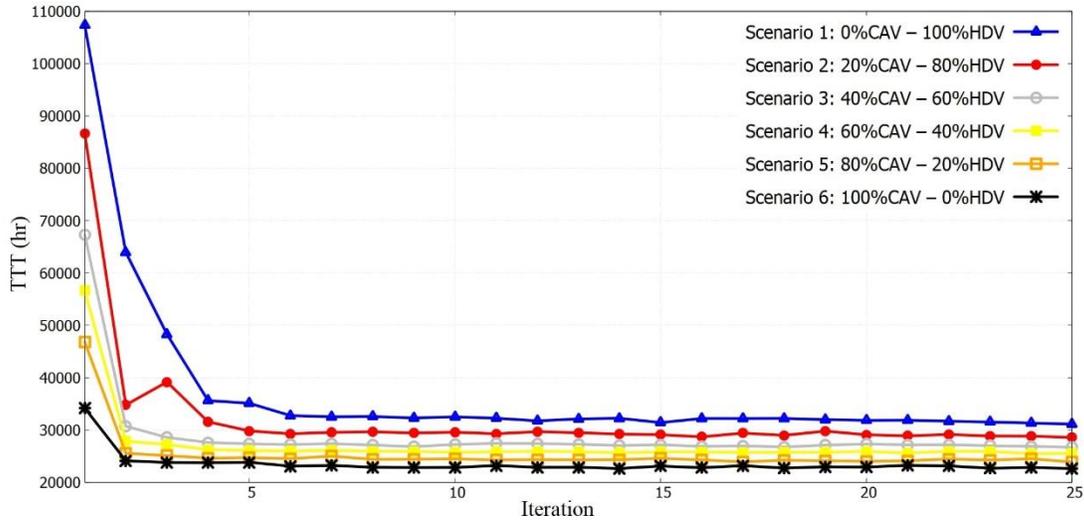

Figure 6: Convergence patterns for Sioux Falls network

The simulation results (Table 6) indicate that in the full automation scenario (scenario 6), the TTT has the minimum value in comparison with other PRs of CAVs. The percentage of TTT reduction varies between 8.2% to 27.2%. Also, the average speed and distance traveled rise as the PR of CAVs increases.

Table 6: Simulation Results for Sioux Falls Network

| Scenario | Hybrid Gap | Total Travel Time (hr) | Average Speed (km/h) | Average Distance Travelled (km) | TTT improvement (%) |
|---|---|---|---|---|---|
| 1 | 4.44 | 31111.94 | 43.56 | 8.2 | - |
| 2 | 3.41 | 28566.56 | 45.57 | 8.3 | 8.2 |
| 3 | 3.11 | 26670.89 | 46.87 | 8.3 | 14.3 |
| 4 | 2.83 | 25475.60 | 47.95 | 8.3 | 18.1 |
| 5 | 2.51 | 23862.69 | 49.17 | 8.4 | 23.3 |
| 6 | 2.12 | 22636.86 | 49.93 | 8.4 | 27.2 |

Figure 7 shows the volume on the Sioux Falls network for different PRs of CAVs, categorized based on both colors and width, where thicker links indicate higher volume. Focusing on the difference between the traffic volumes, this figure shows that vehicles are distributed among the entire network in scenarios with high penetration rate of CAVs (avoiding selfish routing). They select unused links, minimizing the travel time in the entire network. Also, in high PRs of CAVs links capacity increases which can leads to higher number of links with high volume. Because the same links can service more vehicles in less time in comparison with scenarios with less PRs of CAVs.

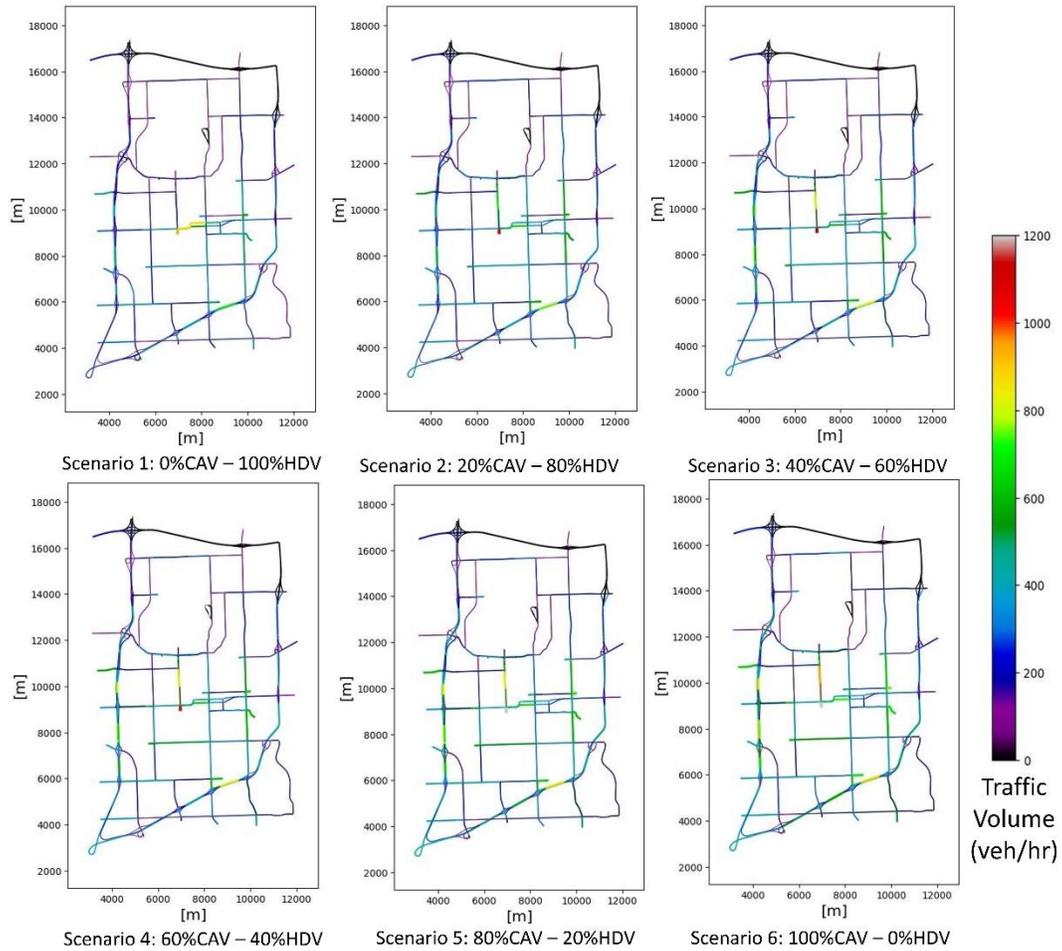

Figure 7: Traffic Volume of Sioux Falls Network in different Penetration Rate of CAVs

Similar to the previous case study to investigate the isolated impact of each CAV-related driving and route choice behavior, several other scenarios are performed and the results are illustrated in Figure 8. The best improvement in TTT in different settings of CAVs impact ranges between 10.1% for the isolate impact of CAV-specific route choice behavior, 20.2% for the isolate impact of CAV-specific driving behavior, and 27.2% for the combined impact of CAV-specific route choice and driving behavior.

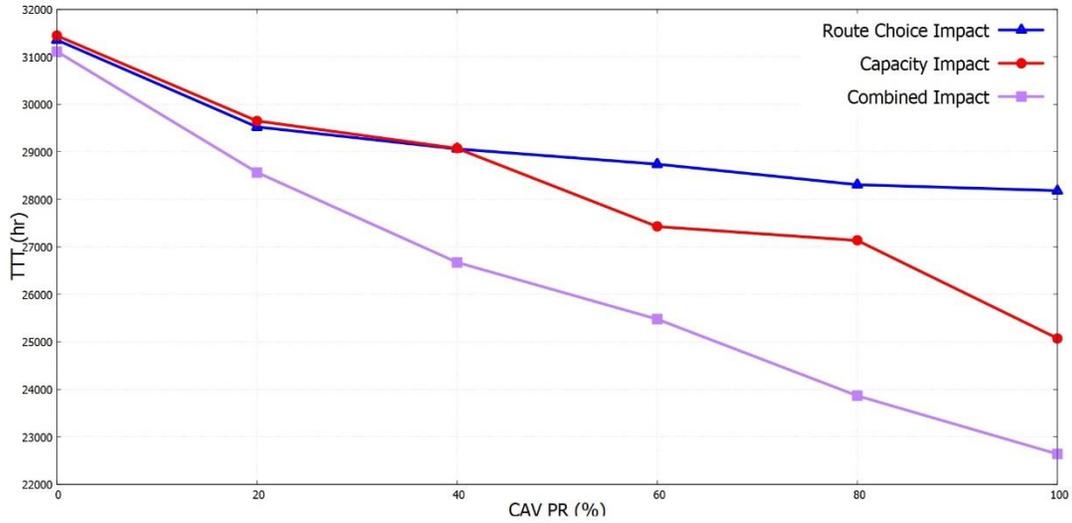

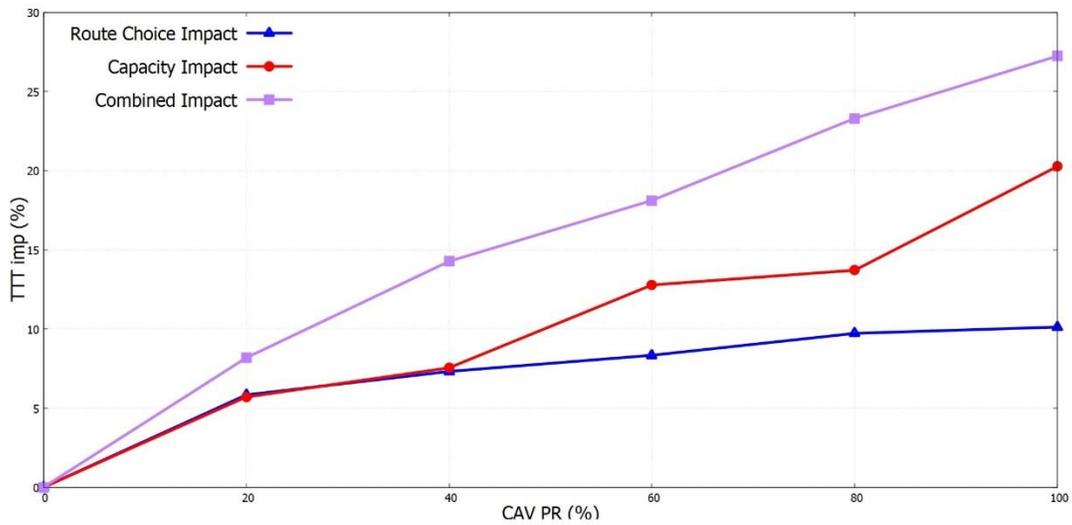

Figure 8: Impact of CAVs on TTT for Different Setting of Route Choice and Driving Behavior (Sioux Falls Network)

## 5- Conclusion

CAVs will revolutionize the transportation sector soon. One of the potential capabilities of CAVs is that their route choice behavior will be different from HDVs. The reason behind this is because a traffic management center can control this type of vehicle. In the near future, we may witness a multiclass equilibrium where different road users with different behaviors coexist simultaneously. Therefore, solving the traffic assignment problem for various PRs of CAVs and HDVs is highly important. This paper presented a solution framework for the Multiclass Simulation-based Traffic Assignment Problem for the Mixed traffic flow of CAVs and HDVs (MS-TAP-M). This problem is defined as the assignment of two classes of vehicles that have different route choices and driving behaviors from each other. It is assumed that CAVs follow the system optimal principle with rerouting capability while HDVs follow user equilibrium routines. To address the impacts of CAVs on road capacity, the modified parameters of the car-following/lane-changing models (microscale) and the queuing model (meso scale) are implemented. The solution framework is an iterative process that includes path selection and dynamic network loading. It starts from an initial solution and updates the path flow distribution for each iteration based on a path-swapping algorithm. At the end of each iteration, a convergence criterion is calculated to check the termination of the algorithm.

To the best of the authors' knowledge, the proposed solution framework for the MS-TAP-M is the first open-source algorithm that can be performed in both micro and meso scale simulation and which considers the rerouting capability of CAVs on top of their different route choice and driving behavior. This solution framework is named "duaIterateMix" and is freely available under the EPLv2 license on GitHub https://github.com/eclipse/sumo/blob/main/tools/assign/duaIterateMix.py. Researchers and decision-makers can freely use this tool to investigate the impacts of CAVs on the road network in various scenarios. duaIterateMix support different assumptions on CAVs' route choice and driving behavior.

In this study, duaIterateMix has been tested on two case studies (both micro and meso scale). Several different settings of CAV-specific route choice and driving behavior have been simulated. The results suggest that the higher the PR of CAVs, the lower TTT of vehicles. In both case studies, the highest improvement in TTT belongs to the time when CAVs have both different route choice and driving behavior (48.9% for the Random network and 27.2% for the Sioux Falls network in TTT improvement). The second significant impact on TTT belongs to the scenarios when CAVs have only different driving behavior (32.6% for the Random network and 20.2% for the Sioux Falls network in TTT improvement). In addition, when CAVs have only different route choice behavior, the least significant impact on TTT occurs (15.8% for the Random network and 10.1% for the Sioux Falls network in TTT improvement). These types of investigations can help researchers and decision-makers to have a more in-depth view of CAVs various potential capabilities. It is recommended to apply the proposed framework on larger networks to check its performance for future works.

## 6- Author Contributions

Behzad Bamdad Mehrabani: Conceptualization, Methodology, Software, Investigation, Data Curation, Writing – original draft, Visualization. Jakob Erdmann: Methodology, Software, Validation, Formal Analysis, Data Curation, Writing – review & editing. Luca Sgambi: Conceptualization, Validation, Writing – review & editing, Supervision, Project Administration, Funding Acquisition. Seyedehsan Seyedabrishami: Methodology, Validation, Writing – review & editing. Maaike Snelder: Conceptualization, Methodology, Validation, Analysis and interpretation of results, Writing – review & editing. All authors reviewed the results and approved the final version of the manuscript.

## 7- Funding

The corresponding author was supported by the Université catholique de Louvain under the ''Fonds Speciaux de Recherche'' and the "Erasmus +" programs.

# 8- References